\documentclass[10pt,journal,compsoc]{IEEEtran}
\ifCLASSOPTIONcompsoc
  \usepackage[nocompress]{cite}
\else
  \usepackage{cite}
\fi
\ifCLASSINFOpdf
\usepackage[pdftex]{graphicx}

\else
\fi
\usepackage{amsmath,amssymb}
\usepackage{graphics, array, url, hyperref, epsfig, amsmath, amssymb}
\usepackage{graphicx,cite,color}
\usepackage{subfigure}
\usepackage{latexsym,amsfonts}
\usepackage{epstopdf}
\usepackage{fancybox}
\usepackage{multirow}
\usepackage{subfigure}
\usepackage{tabularx}
\usepackage{cite}
\usepackage{CJK}
\usepackage{color}
\usepackage{indentfirst}
\usepackage[justification=centering]{caption}

\begin{document}

%
\title{LRCoin: Leakage-resilient Cryptocurrency Based on Bitcoin for Data Trading in IoT}
%
%
%

\author{Yong~Yu$^{\ast}$, 
        Yujie Ding, Yanqi Zhao,~Yannan~Li$^{\ast}$\thanks{$^\ast$ Corresponding Author}, Yi Zhao,
       Xiaojiang~Du, 
       Mohsen Guizani 

\IEEEcompsocitemizethanks{\IEEEcompsocthanksitem Yong Yu, Yujie Ding, Yanqi Zhao, Yannan Li and Yi Zhao are with School of Computer Science, Shaanxi Normal University, Xi'an, 710062, China.\protect\\
E-mail: yuyong@snnu.edu.cn, liyannan2016@163.com
%

\IEEEcompsocthanksitem Xiaojiang Du is with Dept. of Computer and Information Sciences, Temple University, Philadelphia PA 19122, USA.
%

\IEEEcompsocthanksitem Mohsen Guizani is with Dept. of Electrical and Computer Engineering, University of Idaho, Moscow, Idaho, USA. 
}
\thanks{Manuscript received Feb 08, 2018; revised August 25, 2018.}
}

\markboth{Journal of \LaTeX\ Class Files,~Vol.~14, No.~8, August~2015}%
{Shell \MakeLowercase{\textit{et al.}}: Bare Demo of IEEEtran.cls for Computer Society Journals}
%



\IEEEtitleabstractindextext{%
\begin{abstract}

Currently, the number of Internet of Thing (IoT) devices making up the IoT is more than 11 billion and this number has been continuously increasing. The prevalence of these devices leads to an emerging IoT business model called Device-as-a-service(DaaS), which enables sensor devices to collect data disseminated to all interested devices. The devices sharing data with other devices could receive some financial reward such as Bitcoin. However, side-channel attacks, which aim to exploit some information leaked from the IoT devices during data trade execution, are possible since most of the IoT devices are vulnerable to be hacked or compromised. Thus, it is challenging to securely realize data trading in IoT environment due to the information leakage such as leaking the private key for signing a Bitcoin transaction in Bitcoin system. In this paper, we propose LRCoin, a kind of leakage-resilient cryptocurrency based on bitcoin in which the signature algorithm used for authenticating bitcoin transactions is leakage-resilient. LRCoin is suitable for the scenarios where information leakage is inevitable such as IoT applications. Our core contribution is proposing an efficient bilinear-based continual-leakage-resilient ECDSA signature. We prove the proposed signature algorithm is unforgeable against adaptively chosen messages attack in the generic bilinear group model under the continual leakage setting. Both the theoretical analysis and the implementation demonstrate the practicability of the proposed scheme.

%
%
%
%
%

\end{abstract}

}

\maketitle

\IEEEdisplaynontitleabstractindextext

%
\IEEEpeerreviewmaketitle

\IEEEraisesectionheading{\section{Introduction}\label{sec:introduction}}

\IEEEPARstart{T}{HE} Internet of Thing (IoT) has emerged as an area of incredible potential and impact. According to Gartner Inc.\cite{gartner}, more than 11 billion IoT devices have been connected to the IoT network in 2018. The number of these devices is continuously increasing and is expected to be 20 billion by 2020. The application of IoT pervades everywhere from smart home, smart cities, manufacturing, commerce, education to supply chain, logistics and almost anything we can imagine \cite{cisco} -\cite{Hao2016}. The opportunities presented by IoT raise an emerging IoT business model pattern Device-as-a-service(DaaS). We can employ the sensor devices to collect data which would be vended to all interested users or devices. For example, the owner of personal weather station not only uses the IoT devices to control his household heating, but share the data to neighborhood for obtaining financial incentives.

 However, most of these IoT devices are easy to be hacked or compromised by various cyber attacks such as a side channel attack. Due to this attack, the secret key in IoT devices might be leaked and then adversaries could successfully forge valid signatures to transfer bitcoins from those devices to other accounts. Actually, ten thefts of over 10,000 BTC each and more than 34 crimes of stealing accidents over 1000 BTC each since 2011 happened \cite{Gennaro2016}. It was reported by Kaspersky labs that about one million infections per month of malware designed to search for secret keys and steal bitcoins were detected\cite{Gennaro2016}. Dubbed Satori IoT Botnet exploits zero-day to zombify Huawei Routers and was found infecting more than 200,000 IP addresses in just 12 hours \cite{check}. Cisco's Talos cyber intelligence units discovered an advanced piece of IoT botnet malware, dubbed VPNFilter, that was designed with versatile capabilities to gather intelligence, interfere with internet communications, as well as conduct destructive cyber attack operations. The malware infected over 500,000 devices in at least 54 countries, most of which are small and home offices routers and internet-connected storage devices from Linksys, MikroTik, NETGEAR, and TP-Link. Some network-attached storage devices were targeted by the malware as well \cite{thehackernews}. To sum up, the security of IoT devices was degraded due to a variety of factors, and the increasing IoT applications based on Bitcoin leads to an urgent need for more secure bitcoin transactions\cite{LI}.

 Blockchain was firstly introduced by Satoshi Nakamoto in Bitcoin white paper \cite{Nakamoto}. A blockchain is a hash-based data structure. Each block has a block header, a hash pointer to the previous block and a Merkle hash tree (MHT) that digests of transactions in the block. A blockchain makes use of two well-known cryptographic techniques, namely digital signatures and hash functions. A digital signature is employed to provide the integrity, non-repudiation and authentication of bitcoin transactions. A hash function is used to compute a hash value of the previous block and make the blocks as a chain. The decentralization of blockchain benefits IoT in many applications such as access control to data \cite{access}, data trading  \cite{kay}, and key management \cite{Du1}-\cite{tifs} in IoT etc.

\textbf{{Related Work.}}
Noyen et al. \cite{kay} discussed how sensing-as-a-service can benefit from Bitcoin and described the process of exchanging data for cash via Bitcoin. Zhou et al.\cite{zhou} presented distributed data vending on blockchain by combining data embedding and similarity learning. This approach brings the trade-off between the effectiveness of data retrieval and leakage risk from indexing the data. Leiba et al.\cite{leiba} used blockchain as a decentralized IoT software update delivery network in which participating nodes as distributors are compensated by vendors with digital currency for delivering updates to devices. Delgado-Segura et al \cite{segura} introduced a fair protocol for data trading based on Bitcoin script language and double ECDSA. In practically, the script language operator was disabled for Bitcoin transaction. Kopp et al. \cite{koppercoin} presented KopperCoin, a distributed file storage system with financial incentives. Later, Kopp et al. proposed privacy-preserving distributed file storage system with financial incentives\cite{koppercoinpp}, which takes advantages of ring signatures and one-time addresses to realize a privacy-preserving payment mechanism. However, these solutions provide no confidentiality and reliability of data in the context of side channel attacks.

\textbf{{Our Contributions.}} Blockchain is subject to side-channel attacks due to the openness of its deployment. As a consequence, information leakage especially secret key leakage is possible in various IoT applications. In this paper, we propose a new kind of cryptocurrency named LRCoin which is secure even part of the signing key of a user is exposed. LRCoin can be used in the applications where secret key leakage is inevitable such as the payment of data trading in IoT. We propose a concrete construction of an efficient bilinear pairing-based continual leakage-resilient ECDSA signature algorithm as the building block for signing transactions in LRCoin. The proposed signature algorithm is proven unforgeable against adaptively chosen messages attack in the presence of continual leakage setting. The security proof is conducted in the generic bilinear group model to bypass the impossible results that achieving continual leakage-resilience cryptographic protocols whose secret key is uniquely determined by the corresponding public key. We also implement the proposed signature algorithm on laptops and phones respectively, which demonstrates that its efficiency is comparable with that of the original ECDSA signature.

\section{Data Trading Model in IoT}\label{sec:tranding}
\begin{figure}[h]
  \centering
  \includegraphics[width=0.5\textwidth]{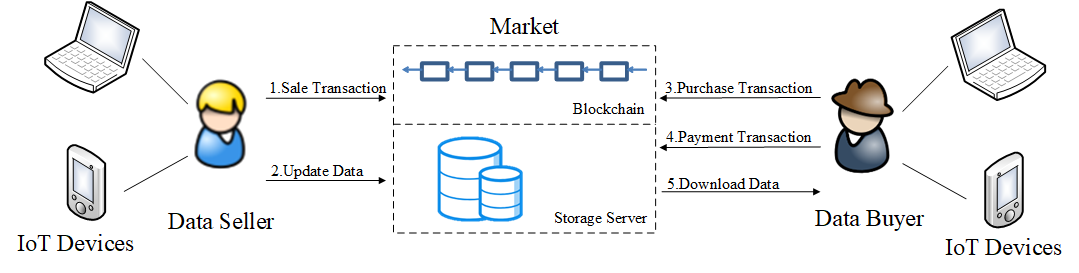}\\
  \caption{Data Trading Model in IoT}\label{fig:ress}
\end{figure}

Data trading \cite{NZ},\cite{survey} is the exchange of bitcoin for data collected by IoT devices between data seller and data buyer. The market is to establish a platform for data seller to use blockchains as infrastructures to sell the data. The data buyer can retrieve data from blockchain and complement the payment. In this section, we introduce the data trading model in IoT and the key components.
The participants involved in data trading model in IoT include Data Sellers, Trade Market (Blockchain Network, Storage Server), Data Buyers, as shown in Fig.\ref{fig:ress}.

{Data seller:} A data seller is a user who owns the data and wants to sell the data.

{Trade market:} The trade market is composed of blockchain network and storage server. The blockchain network provides data sellers and data buyers a trading platform. The storage server provides data sellers and data buyers an intermediate facility to upload data and download data respectively.

{Data buyer:} A data buyer submits Purchase Transactions to the trade market to match the corresponding Sale Transactions. A data buyer generates a payment transaction for data she wants to buy and downloads the data from the storage sever.


{Sale Transaction:} A data seller constructs a Sale Transaction with the topic that he has some data to sell and then broadcasts the transaction to the trade market and uploads the data to the storage server.
A sale transaction consists of the topic of the data, the intended price of selling the data etc.


{Purchase Transaction:} A data buyer constructs a purchase transaction with the topic of the data that she wants to buy and then broadcasts the transaction to data market. A purchase transaction consists of the topic of the data, the intended price of buying the data etc.

\section{Preliminaries}\label{sec:secmodel}

In this section, we recall some preliminaries used in this paper, including bilinear maps, leakage-resilient models and the security model for continual leakage-resilient digital signatures.

\subsection{Bilinear Maps}

Let $G=<g>,G_T=<g_T>$ be two multiplicative cyclic groups of prime order $p$ with $k$ bits. A map\cite{boneh2001} $e:G\times G\rightarrow G_T$ is called a bilinear map if the follow conditions holds.

  \textbf{Bilinearity.}  For all $u,v\in G$ and $a,b\in Z_p$, $e(u^a,v^b)=e(u,v)^{ab}$.

  \textbf{Non-Degeneracy.} $e(g,g)\neq 1_{G_T}$, the identity element of $G_T$.

  \textbf{Efficient Computation.} $e(u,v)$ can be computed in polynomial time.

  \subsection{Stateful Signatures}
In order to achieve continual leakage resilience when a significant bits of secret key are leaked during each round of computations, it is necessary to let the secret key stateful. That is, the secret key must be refreshed after each round of signature computation. Otherwise, the secret key would eventually be completely exposed. Galindo and Vivek \cite{Galindo2013} suggested to split the secret key into two parts and reserve them on two distinct parts of a device memory. Specifically, they divide a signature computation into two steps. In each step, the memory in use is divided into two parts called the active part and the passive part. The active part of the memory is the memory being accessed by the computation while other parts of the memory are the passive part. It is assumed that information leakage is only possible in the active part at any specified time.

There are four polynomial-time algorithms namely $KeyGen,Sign_1,Sign_2,Verify$ in a stateful signature scheme. Different from the key generation of traditional digital signatures which generates a single secret key, the key generation algorithm in a stateful signature outputs two initial secret states $(S_0,S_0')$. The two signing algorithms $Sign_1,Sign_2$ are executed sequentially to generate a signature on a message $m$. A bit more specific, the i-{th} round of the signature computation is performed as follows.
\begin{eqnarray}
           Sign_{1}(S_{i-1}, m_i, r_i) \rightarrow (S_i,w_i),\nonumber\\
           Sign_{2}(S_{i-1}', w_i, r_i') \rightarrow (S_i',\sigma_i),\nonumber
\end{eqnarray}
where $r_i$ and $r_i'$ denote the random values used in $Sign_1$ and $Sign_2$ respectively. $w_i$ denotes the state information delivered to $Sign_2$ by $Sign_1$. Then the secret state is updated from $(S_{i-1},S_{i-1}')$ to $(S_i,S_i')$. The signature of $m_i$ is $\sigma_i$.

The formal definition of a stateful signature $\Pi=(KeyGen,Sign_1,Sign_2,Verify)$ is as follows.
\begin{itemize}
  \item \textbf{KeyGen($k$):} On input a security parameter $k$, it outputs a key pair $(pk,(S_0,S_0'))$ where $pk$ is public key and $(S_0,S_0')$ are two shares of the secret key.
  \item \textbf{Sign$_{1}$($S_{i-1},m_i$):} On input the first part of the $(i-1)^{th}$ secret key $S_{i-1}$ and message $m_i$, it selects a randomness $r_i$ and updates $S_{i-1}$ into $S_i$ and computes the state information $w_i$, which would be passed onto \textbf{Sign}$_{2}$.
  \item \textbf{Sign$_{2}$($S_{i-1}',w_i$):} On input the second part of the $(i-1)^{th}$ secret key $S_{i-1}'$ and state information $w_i$, it chooses a randomness $r_i'$ and updates $S_{i-1}'$ into $S_i'$ and computes the $i^{th}$ signature $\sigma_i$.
  \item \textbf{Verify($pk,\sigma_i,m_i$)}: On input the public key $pk$, signature $\sigma_i$ and message $m_i$, it outputs a bit $b=1$ meaning valid, or $b=0$ meaning invalid.
\end{itemize}

\subsection{Existential Unforgeability with Leakage}
 A cryptographic primitive is called leakage-resilient if it is secure even the adversary additionally obtains some side-channel information, that is leakage. A variety of leakage models \cite{Malkin2011}\cite{Micali2004}\cite{Alwen2009} have been proposed to model side-channel attacks and formalize the security for diverse cryptographic primitives. In 2004, Micali and Reyzin\cite{Micali2004} gave a leakage model namely ``only computation leaks information (OCLI)", saying that only the secret memory which is involved in computation at that time leaks information. And they noted that the leakage amount during each computation is bounded. Otherwise, the adversary continually obtains leakage from many computations, and finally the adversary is able to obtain the full knowledge of the secret key. In 2009, Akavia et al.\cite{Akavia2009} introduced a general leakage model called ``bounded-memory leakage model". In this model, an adversary is allowed to adaptively choose an efficiently computable a leakage function $f$ and send it to a leakage oracle. The adversary obtains $f(sk)$ from the leakage oracle where $sk$ denotes the secret key of the target user. However, the overall output length of all the leakage functions is bounded by a parameter $\lambda$, which is smaller than the secret key $sk$. But this model does not cover the continuous memory leakage, which could been given rise to due to various side channel attacks. In 2010, Zvika et al.\cite{Brakerski2010} and Dodis et al.\cite{Dodis2010} formalized a ``continual-memory leakage model". This model is similar to the OCLI model except that in this model the leakage is assumed from the entire secret memory whether or not the memory is involved in computation.

Galindo and Vivek \cite{Galindo2013} presented an approach to model the leakage in a signature generation by allowing an adversary $\mathcal A$ to access to a leakage oracle $\Omega_{secret key}^{leak}(\cdot)$. It not only gives $\mathcal A$ signatures of messages choosen by $\mathcal A$ but also allows $\mathcal A$ to obtain leakage from the current signature computation. More precisely, let $\lambda$ be the leakage parameter, and $\mathcal A$ is allowed to adaptively select two efficiently computed leakage functions $f_i()\to \{0,1\}^{\lambda}$ and $h_i\to \{0,1\}^{\lambda}$ during every round of signature generation. Specifically, the inputs of leakage functions $f_i$ and $h_i$ are a part of the secret key respectively, and the outputs of leakage are denoted as $\Lambda_i=f_i(\cdot), \Lambda_i'=h_i(\cdot).$ Galindo and Vivek noted that $\mathcal A$ can determine $h_i$ after seeing $\Lambda_i$. But for simplicity, they only define the leakage model in which $f_i$ and $h_i$ are specified along with the message $m_i$ when they are sent to the leakage oracles.

The property of unforgeability of a stateful signature scheme $\Pi=(KeyGen,Sign_1,Sign_2,Verify)$ with continual leakage is defined by the following game between a challenger $\mathcal C$ and an adversary $\mathcal A$.

%

\begin{itemize}
	\item \textbf{Setup}. The challenger $\mathcal C$ runs the key generation algorithm KeyGen$(1^k)$ to obtain a public pair key $pk$ and the initial secret key $(S_0,S_0'))$. $pk$ is given to $\mathcal A$. $\mathcal C$ sets a counter $i=1$ and a set of $\omega=\oslash$ where $i$ denotes the $i^{th}$ round of the signature query and $\omega$ is the set of messages which have been signed by querying the Sign-Leak oracle below.
	\item \textbf{Sign-Leak Queries}. Given with the public parameter $pk$ the adversaries $\mathcal A$ can query a Sign-Leak Oracle $A^{\Omega_{S_{i-1},S_{i-1}'}^{leak}(m_i,f_i,h_i)}(pk)$ at most $q$ numbers of signatures of messages $(m_1,\ m_2,\ ... \ m_i) \in [0,1]^*$ adaptively chosen by $\mathcal A$ $(i<q)$. When $\mathcal A$ queries the Sign-Leak Oracle, If $\lvert f_i \rvert \neq \lambda$ or $\lvert h_i \rvert \neq \lambda$ the oracle would return $\perp$ and then abort. Otherwise the oracle would response $\mathcal A$ with a signature $\sigma_i$ by computing ${Sign}_{1}(S_{i-1},m_i) \stackrel{r_i}{\to} (S_i,w_i)$ and ${Sign}_{2}(S_{i-1}',w_i) \stackrel{r_i'}{\to} (S_i',\sigma_i)$.
       During each such signature computation, the adversaries $\mathcal A$ could also get some knowledge about the internal secret key from the leakage functions $f_i$ and $h_i$ functioning by $\Lambda_i = f_i(S_{i-1},r_i)$ and $\Lambda_i' = h_i(S_{i-1}',r_i', w_i)$.
       After each such query, the counter $i$ is increased to $i+1$ and the messages set is enlarged by $\omega \bigcup m_i$. Eventually the Sign-Leak Oracle returns $(\sigma_i,\ \Lambda_i,\ \Lambda_i')$ to the adversary $\mathcal A$.
   \item \textbf{Output.} Finally, $\mathcal A$ gives a pair $(m,\ \sigma)$. If there exists (1) $Verify(pk,m,\sigma)=1$ and (2) $m \notin \omega$ then the experiment returns $b=1$ meaning that $\mathcal A$ has won the game. Otherwise the experiment returns $b=0$ meaning that $\mathcal A$ has failed to forge a signature.
\end{itemize}

 We define $Pr^{forge}_{\mathcal A}$ as the probability of $\mathcal A$ wins in the above game. The probability $Pr^{forge}_{\mathcal A}$ is taken over the coin tosses of $\mathcal A$ and $KeyGen$.

 \textbf{Definition 1.}  A signature scheme $\Pi$ is ($\epsilon, \tau, q$)-existentially unforgeable under adaptively chosen message attacks with continual leakage if for all ($\epsilon, \tau, q$)-adversaries $\mathcal A$ where $\tau$ is the most running time of $\mathcal A$ and $Pr^{forge}_{\mathcal A}$ is at least $\epsilon$ and $q$ is the most number of queries to oracle, the probability $Pr(b=1)$ in the experiment Sign-Leak$_{\Pi}(\mathcal A,\ k, \ \lambda)$ is negligible(as a function of the security parameter $k$).

\subsection{Generic Bilinear Group Model}

The ``generic algorithm" was first proposed by Shoup, in which the group elements are encoded as unique binary strings and the special properties of the encodings of the group elements are not exploited. This model was extended to the generic bilinear group \cite{Galindo2013} where a bilinear map is involved.

The specific details about the generic bilinear group model have been formalized by Galindo in\cite{Galindo2013}, where representations of bilinear group elements in $\mathbb{G}$ and $\mathbb{G}_T$ are given by random bijective maps $\sigma: \mathbb{Z}_p\to \Xi$ and $\sigma_T: \mathbb{Z}_p\to \Xi_T$. $\Xi$ and $\Xi_T$ are sets of bit strings respectively.
There are oracles $O$, $O_T$ and $O_e$ that can compute the group operations in $\mathbb{G}$ and $\mathbb{G}_T$ and the evaluation of the bilinear map $e$. These oracles accept the representations in $\Xi$ and $\Xi_T$ as inputs and generate such representations as outputs, which are defined as follows:

$$O(\sigma(a),\sigma(b))=\sigma(a+b\ mod\ p)\nonumber$$
$$\ \ \ \ \ O_T(\sigma_T(a),\sigma_T(b))=\sigma_T(ab\ mod\ p)\nonumber$$
$$O_e(\sigma(a),\sigma(b))=\sigma_T(ab\ mod p)\nonumber$$

The generator $g$ of the group $\mathbb{G}$ satisfies $g=\sigma(1)$ and the generator $g_T$ of $\mathbb{G}_T$ satisfies $g_T=e(g,g)=\sigma_T(1)$. Since the representation of $g$ is public, thus, users can efficiently generate random elements in both $\mathbb{G}$ and $\mathbb{G}_T$.

We extend the generic bilinear group model \cite{Galindo2013} slightly to the asymmetric pairing setting denoted as $e:\mathbb{G}_1\times \mathbb{G}_2\rightarrow G_T$ where $\mathbb{G}_1\neq \mathbb{G}_2$. Representations of group elements in $\mathbb{G}_1$, $\mathbb{G}_2$ and $\mathbb{G}_T$ are given by random bijective maps $\sigma_1: \mathbb{Z}_p\to \Xi_1$, $\sigma_2: \mathbb{Z}_p\to \Xi_2$ and $\sigma_T: \mathbb{Z}_p\to \Xi_T$, respectively. Moreover, the generator $P_1$ of $\mathbb{G}_1$ satisfies $P_1=\sigma_1(1)$, and the generator $P_2$ of $\mathbb{G}_2$ satisfies $P_2=\sigma_2(1)$, and the generator $P_T$ of $\mathbb{G}_T$ satisfies $P_T=e(P_2,P_1)=\sigma_T(1)$. The adversary $\mathcal A$ has accesses to these generic group oracles, namely $O_1$, $O_2$ and $O_T$. Let $\tau_1$, $\tau_2$ and $\tau_T$ be the number of queries of $\mathcal A$ to group oracles $O_1, O_2$ and $O_T$.
We define the query form of $\mathcal A$ and the response form of generic group oracles as follows. Let the inputs of the $i^{th}$ query of $\mathcal A$ to $O_T$ be of the form $(X_i,Y_i)$, where $X_i$ and $Y_i$ are representations in $\Xi_T$. And let the response of the generic group oracle $O_T$ to the $i^{th}$ query be the representation $Z_i$, such that $Z_i=X_i \times Y_i$. For convenience, all the above mentioned representations, including inputs and outputs are denoted as $R_1,R_2,...,R_{3\tau_T}$, meaning that $(X_1,Y_1,Z_1,X_2,Y_2,Z_2...)=(R_1,R_2,R_3,R_4,R_5,R_6,...)$. The query form of $\mathcal A$ to $O_1, O_2$ and their response form are similar to $O_T$, except the response of $O_1$ and $O_2$ is $Z_i=X_i+Y_i$. Three tables $\mathcal{L_T}$, $\mathcal{L}_1$ and $\mathcal{L}_2$ defined below are maintained to reserve these group element representations obtained from $O_1$, $O_2$ and $O_T$.
\begin{eqnarray}
\mathcal{L}_T&=&\{R_1,R_2,R_3,R_4,R_5,R_6,...R_{3i}:1\le i\le \tau_T\} \nonumber \\
\mathcal{L}_1&=&\{R_1,R_2,R_3,R_4,R_5,R_6,...R_{3i}:1\le i\le \tau_1\} \nonumber \\
\mathcal{L}_2&=&\{R_1,R_2,R_3,R_4,R_5,R_6,...R_{3i}:1\le i\le \tau_2\} \nonumber
\end{eqnarray}

Since $\mathcal A$ can also query group oracles with representations not previously appeared in the above tables, called independent representations, we introduce another three tables in order to maintain the consistences of the representations of the group elements.
\begin{eqnarray}
\mathcal{Q}_T&=&\{Q_1,Q_2,Q_3,Q_4,Q_5,Q_6,...Q_t:1\le t\le 2\tau_T \} \nonumber \\
\mathcal{Q}_1&=&\{Q_1,Q_2,Q_3,Q_4,Q_5,Q_6,...Q_t:1\le t\le 2\tau_1 \} \nonumber \\
\mathcal{Q}_2&=&\{Q_1,Q_2,Q_3,Q_4,Q_5,Q_6,...Q_t:1\le t\le 2\tau_2 \} \nonumber
\end{eqnarray}
where $Q_1,Q_2,...Q_t$ are received by generic bilinear group oracles in order. It is clear that if the total query number to oracle $O_T$ is $\tau_T$, then $t \le 2\tau_T$ because the number of independent inputs is at most twice of the number of queries. The same conclusions apply to oracles $O_1$ and $O_2$.

\textbf{Lemma 1.} An observer of the interactions between $\mathcal A$ and generic bilinear group oracles can determine, for each representations, a sequence of integers, namely $\bf a_i$$=(a_{i1},a_{i2},...,a_{it})$ with the property that $R_{i}=\sum_{j=1}^t a_{ij}Q_j$ in group $\mathbb{G}_1$ and $\mathbb{G}_2$ or $R_{i}=\prod_{j=1}^t Q_{j}^{aij}$ in group $\mathbb{G}_T$. We call $\bf a_i$ the combination of $R_i$ in generic bilinear group $\mathbb{G}_1$, $\mathbb{G}_2$ and $\mathbb{G}_T$.

Brown \cite{Brown2000} proved this lemma in a single group $\mathbb{G}$, and it holds naturally in $\mathbb{G}_1, \mathbb{G}_2$ and $\mathbb{G}_T$. Thus, lemma 1 holds.

%
%
 If there exist two different combinations $\bf a_{j}$$ \neq \bf a_{k}$ satisfying $R_{j} = R_{k}$, we call that a collision appeared in tables $\mathcal{L}_1$, $\mathcal{L}_2$ and $\mathcal{L}_T$. When a collision is found in $\mathcal{L}_T$, the observer can conclude
\begin{eqnarray}
\bf u^{\bf a_{j}- \bf a_{k}}=1 \quad mod \quad p
\end{eqnarray}
Similarly, when a collision appears in $\mathcal{L}_1$ or $\mathcal {L}_2$, the observer has
\begin{eqnarray}
 (\bf a_{j}- \bf a_{k})\cdot \bf u=0 \quad mod \quad p
\end{eqnarray}
 With these equations, the observer is able to infer some knowledge of the set $\bf u$. Let $r_{i} \in \mathbb{Z}_p$ be the unique unknown value such that $\sigma_T(r_{i})=R_{i}$, then the observer can obtain some information $r_i$ where $r_{i}=\bf a_{i}$$ \cdot \bf u$ in $\mathbb{G}_1$ and $\mathbb{G}_2$ or $r_{i}=$$ \bf u$$^{\bf a_i}$ in $\mathbb{G}_T$.

\textbf{Lemma 2 \cite{Brown2000}.} Suppose there are at most $m$ oracle queries, the probability of a collision occurring in a generic bilinear group for $Z_p$ is at most $3 \binom{m+1}{2}/p$.

\section{Our Construction}\label{sec:construction}

\subsection{Basic idea}
To make the original ECDSA signature scheme continual-leakage-resilient, we apply the techniques due to Galindo et al \cite{Galindo2013}. That is, instead of managing a single secret key, the secret key is divided into two shares which are stored in different parts of the memory. The signing algorithm is divided into two steps as well. We deal with the continual leakage of the secret key by refreshing the two shares of the secret key after each signature round regularly to keep the internal secret key stateful.


\subsection{Detailed construction}

  The details of the proposed leakage-resilient paring-based ECDSA signature are as follows, in which $\mathit{i}$ denotes the $\mathit{i}^{th}$ round of signing.

\textbf{Setup}($k$): On input a security parameter $k$, this algorithm randomly picks a pairing friendly curve $C$ defined on the finite field $\mathbb{Z}_p$ and outputs the corresponding bilinear pairing generic groups $\mathbb{G}_1$, $\mathbb{G}_2$ and $\mathbb{G}_T$, and a bilinear map $e$: $\mathbb{G}_1\times \mathbb{G}_2 = \mathbb{G}_T$. The operations of $\mathbb{G}_1$ and $\mathbb{G}_T$ are denoted as $+$ and $\times$ respectively. This algorithm chooses a base point $P_1$ of $\mathbb{G}_1$ and a base point $P_2$ of $\mathbb{G}_2$, and computes $P_T=e(P_1, P_2)$. $H :\{0,1\}^\star \to \mathbb{Z}_p$ denotes a secure hash function and $\mathit{f}(R)= R \; mod \; p$ denotes an almost invertible reduction function where $R\in \mathbb{G}_T$ defined in\cite{Brown2000}.
The system parameters are denoted as $para=(\mathbb{G}_1,\mathbb{G}_2, \mathbb{G}_T,P_1,P_2,P_T,e,p,C,H,f).$
%

\textbf{KeyGen}($k,para$): On input the security parameter $k$ and the system parameters $para$, this algorithm randomly picks two integers $d, l_0 \in \mathbb{Z}_p$, and computes $S_0 = l_0 P_1$, ${S_0}'=(d-l_0) P_1$ and $Q= P_T^d$. The public key is $pk=(P_1,P_2,P_T,Q)$ while the initial secret key is $sk=(S_0, {S_0}')$.

 \textbf{Sign$_{1}$}($S_{i-1}, m_i$): On input the first part of the $(i-1)^{th}$ secret key, to sign a message $m_i$, this algorithm randomly selects integer $l_i, t_i \in [0,p-1]$ and computes $S_i=S_{i-1}+ l_i P_1$, $R_s= P_T^{t_i}$, $r_s= f(R_s)$, $h_s= H(m_i)$, and $w_i=t_i P_1 + h_s  r_s S_i$.

\textbf{Sign$_{2}$}($S_{i-1}',h_s,r_s,w_i$): On input the second part of the $(i-1)^{th}$ secret key, this algorithm computes $S_i'=S_{i-1}'- l_i  P_1$, and $s_i= w_i + h_s r_s {S_i}'$. The signature on the message $m_i$ is $\sigma_i=(r_s, s_i)$.

\textbf{Verify}($pk, m_i, \sigma_i$): On input a message-signature pair $(m_i, \sigma_i)$ where $\sigma_i=(r_s, s_i)$, this algorithm computes $h_v=H(m_i)$ and
      $R_v=e(s_i,P_2) \times Q^{-(h_v r_s)}$. Output a bit $1$ to indicate the signature is valid if $f(R_v)=r_s$. Otherwise, output $0$ meaning the signature is invalid.

\section{Security Proof}\label{sec:proof}

In this section, we prove the security of our construction against adaptive chosen message attack under continual leakage setting in the generic bilinear group model via a sequence of games. To do this, we firstly prove the security of our scheme denoted as $\Pi$ against ($\Gamma$-time,$0$-query) adversary called a passive adversary under no leakage environment through a game $\mathcal{G}_1$.
Then, we prove security against a ($\Gamma$-time,$q$-query) adversary called an active adversary under no leakage setting through a game $\mathcal{G}_2$. It is clear that the attack power of the adversaries in $\mathcal{G}_2$ is stronger than that of in $\mathcal{G}_1$. More precisely, except the generic group oracles $O_1$, $O_2$ and $O_T$, $\mathcal A$ can also query a signing oracle to obtain polynomial signatures of messages. Next, we prove security against an active adversary under the continual-leakage setting through a game $\mathcal{G}_3$. In this game, the adversary $\mathcal A$ can not only obtain the representations of group elements and signatures of messages but also can get some leaked knowledge of the internal secret key.


\subsection{Proof under no leakage setting}

We firstly provide the unforgeability proof of our scheme against a passive adversary, and then, we give a proof against an active adversary in this part.

\textbf{Against a passive adversary.}

%

\textbf{Theorem 1.} If there exists an $(\epsilon,\tau,0)$-adversary $\mathcal A$ that can forge a valid signature of our scheme $\Pi$ in the generic bilinear group model, then we can construct an $(\epsilon',\tau')$-hash-inverter $I_H$ where
\begin{eqnarray}
\epsilon' \ge \frac{\epsilon - 9 \binom{\tau'+1}{2}/p}{\tau'}.
\end{eqnarray}

$Proof.$ In order to construct a hash inverter $I_H$, we use the adversary $\mathcal A$ as a sub-routine.

In the generic bilinear group model, the functions $\sigma_T$, $\sigma_1$ and $\sigma_2$ behave randomly. Just like hash functions are controlled by the challenger in the random oracle model, $\sigma_T$ is controlled by $I_H$ in the generic bilinear group model.
With this simulation, we describe the response of $I_H$ to group oracle queries as follows.

\textbf{Query to $O_T$:} Upon receiving an independent input $R_{i}=Q_{j}$ for some $i,j$ from $\mathcal A$, the oracle $O_T$ selects a random $u_{j} \in \mathbb{Z}_p \setminus \{r_{1},...r_{i-1}\}$ and sets $\sigma_T(u_{j})=Q_{j}$. $O_T$ adds $Q_{j}$ into the table $\mathcal{Q}_{T}$ and adds $u_{j}$ to $\bf u$. If $O_T$ receives a dependent input $R_{i}=R_{k}$ for some $k < i$, it sets $r_{i}=r_{k}$. Before $O_T$ outputs the response $Z_{3i}$, it first computes $\bf a_{3i}=a_{i}+a_{i+1}$ where $\bf a_{i}$ and $\bf a_{i+1}$ can be determined from $\mathcal A$'s two inputs. Then it computes $r_{3i}=\prod_{j=1}^t u_j^{a_{3i,j}}$. Next, it compares $r_{3i}$ with $\{r_1,...,r_{3i-1}\}$, which are already in the table $\mathcal{L}_T$. If $r_{3i}=r_{k}$ for some $k < 3i$, then $O_T$ responds with $R_{k}$. If $r_{3i} \neq r_{k}$ for all $k < 3i$ then $O_T$ selects randomly $R_{3i} \in \Xi_T \setminus \{R_{1},...,R_{3i-1}\}$ and appends it into the table $\mathcal{L}_T$.

\textbf{Query to $O_1$ and $O_2$:} $\mathcal A$'s queries to generic group oracles $O_1$ and $O_2$ are similar to queries to $O_T$. A difference is that $O_1$ and $O_2$ computes the pre-image as $r_{3i}=$$\bf a_{3i}$$\cdot \bf u$$=\sum _{j=1}^t a_{3i,j} \cdot Q_j$.

\textbf{Description of game $\mathcal{G}_1$:} We describe a reduction game from an $(\epsilon,\tau,0)$-adversary $\mathcal A$ to a $(\epsilon',\tau')$-hash-inverter $I_H$ of the hash function $H$. More precisely, if $\mathcal A$ can forge a signature in probability $\epsilon$, then $I_H$ can invert $H$ in probability $\epsilon'$.

The input to $I_H$ is a random element $h\in[0,p-1]$ as a challenge, and the goal of $I_H$ is to find $M$ such that $H(M)=h$. To find $M$, $I_H$ invokes the adversary $\mathcal A$ and let it interact with a modified simulation of generic group oracle $O_T$. When $\mathcal A$ is invoked, it makes some queries to generic group oracles. In the end, $\mathcal A$ outputs $(M,(r_s,s))$, where $M$ is an arbitrary message and $(r_s,s)$ is a signature for $M$ valid under the public key $Q$. Initially, we assume without loss generality that the first independent representation in table $\mathcal{Q}_1$ is the base point $P_1$ in $\mathbb{G}_1$ and $P_1=Q_1=\sigma_1(1)$. And the first independent representation in table $\mathcal{Q}_2$ is the base point $P_2$ in $\mathbb{G}_2$ and $P_2=Q_1=\sigma_2(1)$. 
In the table $\mathcal{Q}_T$, the first independent representation is the signer's public key $Q$ and $Q=Q_{1}=\sigma_T(X)$ where $X$ denotes the discrete logarithm of the signer's secret key.

To make sure the output message $M$ of $\mathcal{A}$ is the answer of the inverter $I_H$, we introduce the following trick. We conduct a modified simulation of the generic bilinear group oracle $O_T$. Moreover, we ensure that the modified version of the oracle $O_T$, from $\mathcal A$'s perspective, is indistinguishable from the standard version. Therefore, $\mathcal A$ would operate as if it were communicating with a true generic group oracle.

We modify a query from $\mathcal A$ to the generic group oracle $O_T$ as follows. Recall that, all queries of $\mathcal A$ to $O_T$ have an unique combination $\bf a$. We consider that the $i^{th}$ query such that $\bf a_{i+1}$ has the form $(-y,\arrowvert \bf 0)$ where $y \in \mathbb{Z}_p$, and $\bf a_{i+1}$$\ne \bf a$$_l$ for $l\in [1,i]$ and $\bf a_{3i}$ $=\bf a_i$ $+ \bf a_{i+1}$ $\ne \bf a_k$ for $k\in [1,3i-1]$. We call this $i$ $\sl special$. For this special $i$, we modify the way of $O_T$ generating the output $R_{3i}$ as follows. When $I_H$ receives the $i^{th}$ query to $O_T$, say two inputs $(R_i, R_{i+1})$, $O_T$ sets
\begin{eqnarray}
R_{3i}=\mathit{g}(y\cdot h^{-1} \quad mod \quad p),
\end{eqnarray}
where $\mathit{g}$ is the probabilistic inverse of $\mathit{f}$, which exists since $\mathit{f}$ is almost invertible. Moreover, since $h$ is selected randomly and uniformly from $[0,p-1]$, so is $y\cdot h^{-1} \quad mod \quad p$. Thus, the distribution of modified simulation $R_{3i}$ is indistinguishable from the uniform distribution over $\Xi_T$.
As a result, $\mathcal A$ is able to forge a signature $(r_s,s)$ for message $M$ as it normally would with the true simulation.

Upon receiving the forgery $(M,(r_s,s))$ from $\mathcal{A}$, $I_H$ computes $R_v=e(s,P_2)\times Q^{-(H(M)\cdot r_s)}$ using the method of double-and-multiply among point elements in $\mathbb{G}_T$, with additional $n$ queries to the modified simulation of the generic group oracle $O_T$. After the last query, $I_H$ gives the response $R_{3(m+n)}=R_v$, which has a unique combination $\bf a_v$$=\bf a_i$$+\bf a_{i+1}$.

Note that $R_v=e(s,P_2)\times Q^{-(H(M)\cdot r_s)}$, which is assumed to be the $i^{th}$ query of $\mathcal{A}$ to $O_T$ at this moment. Next we present the response of $O_T$ to this query. Intuitively, there are two possibilities for $\bf a_v$. If $e(s,P_2)$ is an independent representation then $\bf a$$_i$ $=(\bf 0$$\arrowvert 1)$ and $\bf a_v$ $=(-H(M)\cdot r_s,0,0,...,1)$. Else if $e(s,P_2)=R_{l}$ for $l< \tau_T$, then $\bf a_v$ $=\bf a$$_{l}$ $+(-H(M)\cdot r_s \arrowvert \bf 0)$ $=(a_{l,1}-H(M)\cdot r_s,a_{l,2},a_{l,3},...,a_{l,t})$. If $\bf a_v$ $\ne \bf a$$_l$ for all $l < 3(m+n)$ then $R_v=R_{3(m+n)}\ne R_l$ for $l<3(m+n)$. 
In this case, $I_H$ selects a representation $R_v$ from $\Xi_T \setminus \{R_{T1},R_{T2},...R_{3(m+n)-1}\}$.  Since $A$'s forgery is valid, so there is $\mathit{f}(R_v)=r_s$. If $R_v$ is selected randomly, the probability of $\mathit{f}(R_v)=r_s$ is at most $3/(p-3(m+n))$. Therefore, we assume $\bf a_{3(m+n)}$ $=\bf a$$_l$ for $l< 3(m+n)$. In this case $e(s,P_2)$ is not an independent input. Because if it was independent, the combination $\bf a_v$ $=(0,H(M)r_s,0,...,1)$, which has one more integers than $ \bf a$$_l$ for $l<3(m+n)$. So $\bf a_i$ $=\bf a_k$ for $k<i$ and $\bf a_v$ $= \bf a$$_l$ $=(a_{k,1}-H(M)\cdot r_s,a_{k,2},a_{k,3},...,a_{k,t})$. Obviously, $\bf a$$_l$ $\ne \bf e_i$, so $R_l$ is not independent. Therefore, $R_l$ first appeared as an output, say $R_l=R_{3g}$, for $g \in [1,l-2]$. As a result $\bf a_l$ $=\bf a$$_{3g}$ $=\bf a_{3(m+n)}$. $I_H$ only modifies the response of one of these queries, denoted as $j$. Moreover this $j^{th}$ query was chosen randomly by $I_H$ before game. So the probability of $j=g$ is at least $1/m$, assuming there has been $m$ queries until now. Because $$R_v=R_{3j}=\mathit{g}(h^{-1}\cdot H(M)\cdot r_s)$$ and $$\mathit{f}(R_v)=h^{-1}\cdot H(M)\cdot r_s=r_s,$$ so $H(M)=h\quad mod\quad p$ and $I_H$ outputs $M$ as the solution to the given hash inverse problem, such that $H(M)=h$.


The probability $\epsilon'$ is bounded as follows. Certainly, we need $\mathcal A$ to succeed, which occurs with probability $\epsilon$. Moreover, the above proof assumes no collision occurs, so the probability of collision occurs must be subtracted, which is at most $9\binom{m+n+1}{2}/p$. What's more, the aforementioned proof requires $j=g$, which occurs with probability at least $1/m$. So overall, $$\epsilon'\ge \frac{\epsilon - 9\binom{m+n+1}{2}/p}{m} \ge \frac{\epsilon - 9\binom{\tau}{2}/p}{m}\ge \frac{\epsilon - 9\binom{\tau}{2}/p}{\tau}.$$ Because $$\frac{\epsilon - 9\binom{\tau}{2}/p}{\tau}\ge \frac{\epsilon - 9\binom{\tau'}{2}/p}{\tau'},$$ we get $$\epsilon'\ge \frac{\epsilon - 9\binom{\tau'}{2}/p}{\tau'}.$$

\textbf{Against an active adversary.}


\textbf{Theorem 2.} If there exists an $(\epsilon,\tau,q)$-adversary $\mathcal A$ that can forge a valid signature of the proposed scheme $\Pi$, then there exists an $(\epsilon',\tau')$-collision-finder $C_H$ where
\begin{eqnarray}
\epsilon' \ge \epsilon - 9 \binom{\tau'}{2}/p
\end{eqnarray}

$Proof.$ The adversary $\mathcal A$ can adaptively choose messages to query signatures to $C_H$. Let $\omega$ be the set of messages queried by $\mathcal A$. For simplicity, we only describe the differences between the proof of Theorem 1 and Theorem 2.

\textbf{Description of game $\mathcal{G}_2$:}

In this game, we aim to construct a collision finder $C_H$ using the adversary $\mathcal A$. Similarly, $\mathcal A$ can interact with a modified simulation of the generic group oracle $O_T$. The difference is that 
instead of choosing a random $j$ and modifying the response of the generic group oracle $O_T$ to the $j^{th}$ query, $C_H$ now, for each output $Z_{3i}$ of the generic group oracle $O_T$, chooses a random message $\tilde{M_i}$, computes $\tilde{h_i}=H(\tilde{M_i})$ and uses $\tilde{h_i}$ and $\mathit{g}$ to generate a random output in the same way as in the proof of Theorem 1. In this proof, $\mathcal A$ can interact with a signing oracle. For each signing query on a message $\hat{M_i}$, $C_H$ selects a random signature $(\hat{r_{si}},\hat{s_i})$ and $\mathcal A$ can query $O_T$ to validate this signature. We stress that $O_T$ must be simulated by $C_H$ in such a way that the queried signatures are valid, using the almost invertibility of $\mathit{f}$. More precisely, when $C_H$ receives a query message $\hat{M_i}$, it will response $\mathcal A$ with a random signature $(\hat{r_{si}},\hat{s_i})$, and at the same time, $C_H$ computes $e(\hat{s_i},P_2)$, $Q^{-H(\hat{M}_i)\cdot \hat{r_{si}}}$ and stores them in the table $\mathcal{L}_T$. Furthermore, $C_H$ sets $R_v=e(\hat{s_i},P_2)\times Q^{-H(\hat{M}_i)\cdot \hat{r_{si}}}=\mathit{g}(\hat{r_{si}})$ and puts it in the table $\mathcal{L}_T$. Then when $\mathcal A$ queries $O_T$ with $R_v$ to validate that signature, $\mathcal A$ will obtain a value previously computed by $C_H$. Obviously, this signature is valid with the almost invertibility of function $\mathit{f}$. 
On the other hand, when $\mathcal A$ queries $O_T$ with special inputs $X_i$ and $Y_i$, $C_H$ selects a random message $\tilde{M_i}$ and responds $A$ with $g(H(\tilde{M_i})^{-1}\cdot y)$ where $y\in [1,p-1]$ is the first integer in combination $\bf a$$_{Y_i}$. Eventually, if $A$ outputs a valid forgery $(r_s,w)$ for $M$, then there exists $\mathit{f}(R_v)=r_s$. We describe the generation of $R_v$ as follows. In the modified simulation, $C_H$ selects a random $\tilde{M}$ and computes $R_v=g(H(\tilde{M})^{-1}\cdot (H(M)\cdot r_s))$. To achieve $\mathit{f}(R_v)=r_s$, there is $H(\tilde{M})= H(M)$. So overall, under no collision environment, there are two possibilities for $\mathcal A$ to forge a valid signature. One is $H(M)=H(\tilde{M})$ and another is $H(M)=H(\hat{M})$. Clearly, by the definition of a forgery, there is $M\ne \hat{M}$. Furthermore, as the distribution of $M$ is uniform and $\tilde{M}$ is selected randomly, so is $M\ne \tilde{M}$. Thus, if $\mathcal A$ succeeds, then $C_H$ can find two messages ($M$ ,$\tilde{M})$ or $(M,\hat{M})$ with the same hash value. The probability $\epsilon'$ is bounded by $\epsilon' \ge \epsilon - 9 \binom{\tau'}{2}/p$.

\subsection{Proof under leakage setting}

In this proof, we further strengthen the attack power of the adversary $\mathcal A$. That is, in addition to group elements, signature queries of adaptively chosen messages, $\mathcal A$ can also obtain leakages of secret keys used in computing those signatures.

\textbf{Theorem 3.} Let $\mathcal A$ be an $(\epsilon,\tau,q)$-adversary that can forge a valid signature of $\Pi$, then we can construct an $(\epsilon',\tau')$-collision-finder $C_H$ where
\begin{eqnarray}
\epsilon' \ge \epsilon - \frac{\tau'^2}{p} 2^{2 \lambda}.
\end{eqnarray}

\textbf{Description of game $\mathcal{G}_3$:}

In game $\mathcal{G}_2$, the advantage of $\mathcal A$ is bounded by its success probability conditioned on the event that no collision has occurred in the lists consisting of elements of $\mathbb{G}_1$, $\mathbb{G}_2$ and $\mathbb{G}_T$. Note that the proof under non-leakage setting in $\mathcal{G}_2$ and leakage setting in $\mathcal{G}_3$ would be the same conditioned on the fact that a collision has not occurred. The reason is that in the event of no collision, in order to forge a valid signature the adversary has to find a collision of the hash function. Hence, the success probability of $\mathcal A$ against scheme $\Pi$ in and not in leakage setting is same as that in the event of no collision.
In the leakage setting, because $\mathcal A$ has access to leakage oracles $\mathit{f}_i(sk)$ and $h_i(sk)$ during the $i^{th}$ signature computation, then in adversary's view the secret key is no longer uniformly distributed.
Thus, the probability that a collision occurs in the leakage setting is increased by a factor of at most $2^{2\lambda}$. Hence $\mathcal A$ can now cause collisions among representations in tables $\mathcal{L}_T$, $\mathcal{L}_1$ and $\mathcal{L}_2$ with the increased probability. The output of each leakage function is at most $\lambda$ bits. So overall, there would be at most $2\lambda$ bits leaked. So in the view of the adversary, the secret key would only have a min-entropy $logp-2\lambda$.

According to Lemma 2, the probability of a collision occurring is increased to
\begin{eqnarray}
\frac {9 \binom{m+1}{2}}{p} 2^{2 \lambda}.
\end{eqnarray}

The probability $\epsilon'$ of $C_H$ in leakage setting is bounded as
\begin{eqnarray}
\epsilon' \ge \epsilon - \frac {9 \binom{m+1}{2}}{p} 2^{2 \lambda}\nonumber \ge \epsilon - \frac{\tau'^2}{p} 2^{2 \lambda}
\end{eqnarray}

\section{Implementations}\label{sec:imp}

In this section, we show the implementations of the proposed leakage-resilient signature algorithm.

\subsection{Environment} We implemented the proposed scheme on a laptop with 4.00 GB RAM, 64-bit Win 7 operating system and a phone with 4.04GB RAM, Android 7.1.1 operating system respectively. The processors of the laptop and phone are Intel(R) Core(TM) i5-2450M CPU @ 2.50GHz and 8 the highest 2.45GHZ respectively. The implementation was conducted with C++ projects and a powerful Miracl library, which realizes kinds of cryptographic operations such as big integers and elliptic curve group operations. And then compile the C++ projects in Visual Studio 2012.

\subsection{Implementation results} In the implementation, we choose the security level as AES-80 bit, and employ Cocks-Pinch curve\footnote{Methords for constructing Pairing-Friendly Elliptic
Curves," http://cacr.uwaterloo.ca/conferences/2006/ecc2006/freeman.pdf} $y^2=x^3-3x+B$, with the embedding degree 2 over a finite field $GF(p)$ where $p=3\pmod 4$.

To test the efficiency of our scheme, we executed four sub-algorithms $Setup$, $KeyGen$, $Sign$ and $Verify$, where $Sign$ includes $Sign1$ and $Sign2$. Each sub-algorithm was run 50 times on the laptop and the phone respectively. The average running time of the sub-algorithms to sign a single message is shown in Table 1. The $Setup$ algorithm generates the public parameters of the construction which on average consumes $23.268$ ms on the laptop and $40.663$ ms on the phone.
For $KeyGen$, it is obvious that the most expensive operations are 2 point multiplications in $\mathbb{G}_1$ and 1 exponentiation in $\mathbb{G}_T$. On average, the $KeyGen$ algorithm costs 1.156 ms on the laptop and 12.380 ms on the phone. In the signature computation, the most expensive operations are 5 point multiplications in $\mathbb{G}_1$, which takes about 11.08 ms on the laptop and 39.09 ms on the phone respectively for the $Sign1$ and $Sign2$ algorithms together. To validate a signature, the most expensive operations in algorithm $Verify$ are 1 bilinear map and 1 exponentiation operation. The $Verify$ takes on average 6.083 ms on the laptop and 14.526 ms on the phone.
\begin{table}
\caption{{\small Time cost of our signature scheme}}\label{tab1}
\begin{center}
\begin{tabular}{|c|c|c|c|c|}
 \hline
 \scriptsize \textbf{Sub-Algorithm} & \scriptsize \textbf{Setup} & \scriptsize \textbf{KeyGen }& \scriptsize \textbf{Sign} & \scriptsize \textbf{Verification}\\
 \hline
 \scriptsize Laptop & \scriptsize 23.268 ms & \scriptsize 1.156 ms & \scriptsize 11.083 ms & \scriptsize 6.083 ms\\
 \hline
 \scriptsize Phone & \scriptsize 40.663 ms & \scriptsize 12.380 ms & \scriptsize 39.083 ms & \scriptsize 14.526 ms\\
 \hline
\end{tabular}
\end{center}
\end{table}

\begin{figure}
	\centering
		\includegraphics[width=0.4\textwidth]{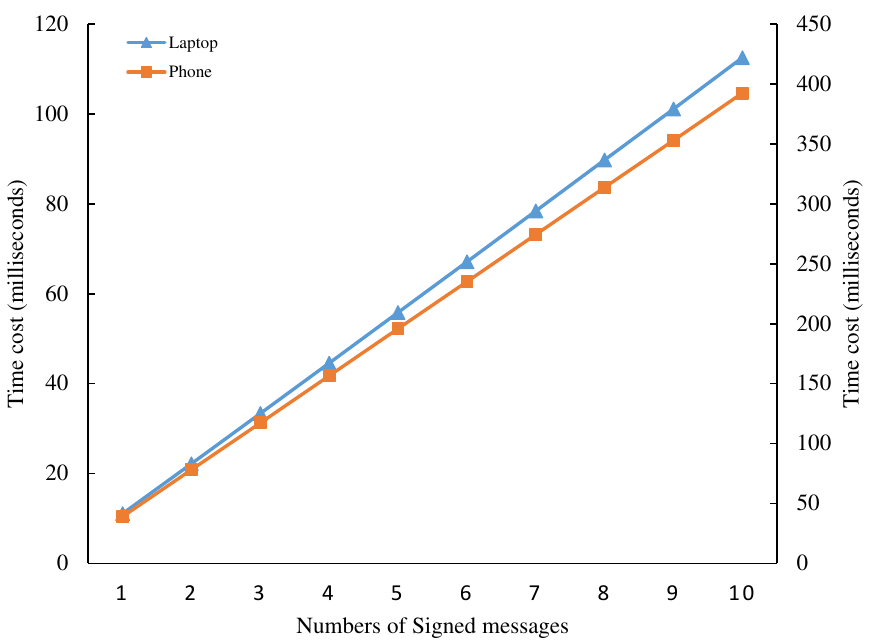}
	\caption{Time cost of \emph{Sign} algorithm}\label{fig:result1}
\end{figure}

\begin{figure}
	\centering
		\includegraphics[width=0.4\textwidth]{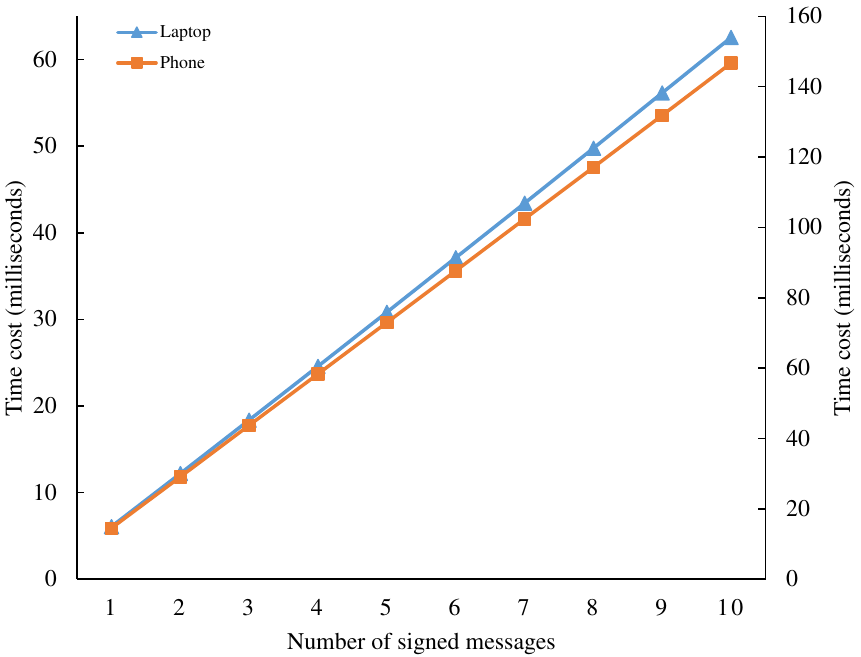}
	\caption{Time cost of \emph{Verify} algorithm}\label{fig:result2}
\end{figure}

In the experiment, the implementation results for $Setup$ and $KeyGen$ algorithms are almost constant both on laptop and phone, that is on average $23.268$ ms and $1.156$ ms on the laptop and $40.663$ ms and $12.380$ on the phone. This is consistent with the empirical analysis since these two algorithms are independent of the signed messages. We also tested the time cost of the $Sign$ and $Verify$ algorithms by increasing the number of signed messages from
1 to 10. The implementation results on the laptop and phone are demonstrated in Fig. \ref{fig:result1} \ref{fig:result2}. As expected, the time cost of both the $Sign$ algorithm and the $Verify$ algorithm increases almost linearly with the increase of the number of signed messages. This is consistent with the theoretical analysis of the proposed scheme too since once a signed message is given, its hash value is determined and all the operations of the $Sign$ and $Verify$ algorithms are determined.

\section{Conclusion}\label{sec:conclusion}

 Information leakage especially secret key leakage is a serious threat in a number of IoT applications. In this paper, we propose LRCoin, a kind of leakage-resilient cryptocurrency based on Bitcoin, secure even part of the signing key is exposed as a helpful supplement of Bitcoin. LRCoin can be applied to the applications where information leakage is inevitable to make the payment of data trading in IoT more reliable. The core of LRCoin is a leakage-resilient digital signature for signing transactions in the network. We propose a concrete construction of an efficient bilinear-based continual-leakage-resilient ECDSA signature algorithm as the building block of LRCoin. We prove the unforgeability of the proposed signature algorithm in the generic group model. The implementations on the labtop and the phone demonstrate the efficiency and the practicability of our proposal.




%

\appendices


\ifCLASSOPTIONcompsoc
\section*{Acknowledgment}
%
This work was supported by National Key R\&D Program of China (2017YFB0802000), National Natural Science Foundation of China (61872229), NSFC Research Fund for International Young Scientists (61750110528), National Cryptography Development Fund during the 13th Five-year Plan Period (MMJJ20170216) and Fundamental Research Funds for the Central Universities (GK201702004).

\ifCLASSOPTIONcaptionsoff
  \newpage
\fi

\end{document}